\documentclass [11pt]{article}
\title{Mechanism for Vanishing Zero Point Energy}
\author{Robert D. Klauber\\1100 University Manor Dr., 38B \\Fairfield, Iowa 52556
\\rklauber(AT)iowatelecom.net, permanent: rklauber(AT)netscape.net}
\date{Orig: Sept 24, 2003; Minor revision: July 19, 2007}

\usepackage{graphicx}
\usepackage{longtable}

\begin{document}

\maketitle

\begin{abstract}

In addition to the standard solutions of the quantum field equations having 
the two forms $e^{\pm (i\omega t\,-\,i{\rm {\bf k}}\cdot {\rm {\bf x}})}$, 
there exist two additional solutions of the form $e^{\pm (i\omega 
t\,+\,\,i{\rm {\bf k}}\cdot {\rm {\bf x}})}$. By incorporating these latter 
solutions, deemed ``supplemental solutions'', into the development of 
quantum field theory, one finds a simple and natural cancellation of terms 
that results in an energy VEV, and a cosmological constant, of zero. This 
fundamental, and previously unrecognized, inherent symmetry in quantum field 
theory appears to provide a resolution of the large vacuum energy problem, 
simply and directly, with no modification or extension to the extant 
mathematics of the theory. In certain scenarios, slight asymmetries could 
give rise to dark energy.

\end{abstract}

\section{Introduction}
\subsection{Cosmological Issues}
As summarized by Peebles and Ratra\cite{Peebles:1}, 
Padmanabhan\cite{Padmanabhan:2003}, and others, there are presently three 
overriding cosmologic issues involving phenomena for which no generally 
accepted theoretical solutions exist: 1) dark matter (non-baryonic, unseen 
``normal'' matter), 2) dark energy (small positive cosmological constant or 
quintessence), and 3) a vanishing sum of zero-point energies. These may be 
related, or they may be unrelated. This article focuses on a possible 
solution to the third of these, though it also suggests a concomitant, 
potentially viable, approach toward resolving the second.

\subsection{Negative Frequencies and Supplemental Solutions}
\label{subsec:negative}
The issue of negative frequency solutions to the relativistic counterparts 
of the Schroedinger equation has a long and variegated history. Such 
solutions constitute a second way to solve the quantum field equations 
beyond those of positive frequency. The question of interpretation of 
negative frequency solutions, one of the most famous in the history of 
science, was answered via second quantization. The solutions to the field 
equations could then be shown to be operators that create and destroy 
states, rather than being states themselves.

By including the negative frequency solutions, the set of solutions to the 
quantum field equations was doubled in size. In this article, it is noted 
that that the number of mathematical solutions to the field equations is 
actually twice that again. The typically unused half of the full set is 
designated herein as the set of ``supplemental solutions'', and probable 
reason is suggested for why it has been all but ignored. It is shown that if 
these supplemental solutions were to generate corresponding vacuum 
fluctuations, and maintained unbroken their natural symmetry with the 
traditional solutions, then the theoretical value for vacuum energy density 
would be zero, and hence, quite close to what is observed. 

\section{The Unused Solutions to the Field Equations}
For simplicity, we consider first the scalar field equation in natural units
\begin{equation}
\label{eq1}
\left( {\partial _\mu \partial ^\mu +\mu ^2} \right)\phi =0
\end{equation}
with traditional eigensolutions
\begin{equation}
\label{eq2}
e^{-ikx}=e^{-i\omega t+i{\rm {\bf k}}\cdot {\rm {\bf x}}},
\end{equation}
\begin{equation}
\label{eq3}
e^{ikx}=e^{i\omega t-i{\rm {\bf k}}\cdot {\rm {\bf x}}}.
\end{equation}
Note, however, that (\ref{eq2}) and (\ref{eq3}) are not the only forms for solutions to (\ref{eq1}). 
The following, obtained by taking $\omega \to -\omega $ in the above 
(consider the symbol $\omega $ as always a positive number), are also 
solutions to (\ref{eq1}).
\begin{equation}
\label{eq4}
e^{i\underline{kx}}=e^{i\omega t+i{\rm {\bf k}}\cdot {\rm {\bf x}}}
\end{equation}
\begin{equation}
\label{eq5}
e^{-i\underline{kx}}=e^{-i\omega t-i{\rm {\bf k}}\cdot {\rm {\bf x}}}
\end{equation}
New notation (see underscoring above) and new terminology are herein 
introduced, wherein (\ref{eq4}) and (\ref{eq5}) are the ``supplemental solutions'' to (\ref{eq1}). 
Upon introduction to supplemental solutions, some at first believe them to 
be already contained in the traditional solutions to the field equations. 
This issue is addressed in Appendix A

Note that the Dirac and Pro\c{c}a equations have such supplemental solutions 
as well. The Dirac equation actually has eight eigenspinor solutions, a set 
of four for $e^{+(i\omega t-i{\rm {\bf k}}\cdot {\rm {\bf x}})}$ and another 
set of four for $e^{-(i\omega t-i{\rm {\bf k}}\cdot {\rm {\bf x}})}$, as one 
would expect. Four (two from the first set and two from the second set) have 
positive energy in the spinor components, and four have negative energy in 
the spinor components. Tradition (and lack of relevance for real particles) 
has led to ignoring of the latter four.

In relativistic quantum mechanics\footnote{ RQM employs the Klein-Gordon, 
Dirac, and Pro\c{c}a equations, but unlike quantum field theory, the 
solutions are states, not operators.} (RQM), the precursor to quantum field 
theory (QFT), (\ref{eq3}) presented a problem as it represented a negative energy 
state. As noted, QFT resolved this by considering solutions having the 
spacetime dependence of (\ref{eq2}) and (\ref{eq3}) as operators, not states. Examination of 
(\ref{eq4}) and (\ref{eq5}) in the context of RQM leads to a similar issue of negative 
energy, as well as an additional one. Momentum direction in (\ref{eq4}) and (\ref{eq5}), if 
they represent states, is in the opposite direction of wave velocity [see 
Appendix A], and hence (\ref{eq4}) and (\ref{eq5}) are unlikely candidates for physical 
particle states. Probably for this and other reasons, solutions of the form 
of (\ref{eq4})and (\ref{eq5}) fell out of favor early on in the development of RQM.

In parallel with the historical development of (\ref{eq2}) and (\ref{eq3}) in QFT, however, 
we can apply second quantization to (\ref{eq4}) and (\ref{eq5}), determine the resultant 
field operator solutions, and see if those solutions might provide anything 
of value in helping to match experiment with theory. 

\section{Supplemental Solutions and QFT}
\label{sec:supplemental}
\subsection{Symmetry of the Lagrangian}
If (\ref{eq4}) and (\ref{eq5}) solve the same free field equations as (\ref{eq2}) and (\ref{eq3}), then it 
follows that the free Lagrangian is symmetric under the transformation 
\textit{$\omega $} $\to $ --\textit{$\omega $}.

\subsection{Klein-Gordon Supplemental Solutions}
Quantum field theory formalism for the supplemental solutions can be 
developed in parallel with the standard approach, using, for scalar fields, 
the following definitions.
\begin{equation}
\label{eq6}
\underline{\phi }\underline{}=\underline{\phi 
}^+\underline{}+\underline{\phi }^-=\sum\limits_{\rm {\bf k}} {\left( 
{\frac{1}{2V\omega _k }} \right)^{1/2}\left\{ {\underline{a}({\rm {\bf 
k}})e^{i\underline{kx}}+\underline{b}^\dag ({\rm {\bf 
k}})e^{-i\underline{kx}}} \right\}} 
\end{equation}
\begin{equation}
\label{eq7}
\underline{}\underline{\phi }^\dag =\underline{\phi }^{\dag 
+}+\underline{\phi }^{\dag -}=\sum\limits_{\rm {\bf k}} {\left( 
{\frac{1}{2V\omega _k }} \right)^{1/2}\left\{ {\underline{b}({\rm {\bf 
k}})e^{i\underline{kx}}+\underline{a}^\dag ({\rm {\bf 
k}})e^{-i\underline{kx}}} \right\}} 
\end{equation}
\begin{equation}
\label{eq8}
{{\cal L}}_{\,\underline{\phi }}^{\,0} =\partial _\mu \underline{\phi }^\dag 
\partial ^\mu \underline{\phi }\underline{}-m^2\underline{\phi }^\dag 
\underline{\phi }\underline{},
\end{equation}
where (\ref{eq8}) is an extra component added to the Lagrangian density for the 
scalar supplemental solutions. The superscript refers to ``free'' 
Lagrangian.

Applying second quantization to the supplemental solutions where
\begin{equation}
\label{eq9}
\left[ {\underline{\phi _r }({\rm {\bf x}},t),\underline{\pi _s }({\rm {\bf 
y}},t)} \right]=\left[ {\underline{\phi _r }({\rm {\bf 
x}},t),\underline{\dot {\phi }_s }^\dag ({\rm {\bf y}},t)} \right]=i\delta 
_{rs} \delta ({\rm {\bf x}}-{\rm {\bf y}}),
\end{equation}
using (\ref{eq6}), (\ref{eq7}) and the definition of the Dirac delta function for finite 
volume
\begin{equation}
\label{eq10}
\delta ({\rm {\bf x}}-{\rm {\bf y}})=\frac{1}{V}\sum\limits_{\rm {\bf k}} 
{e^{-i{\rm {\bf k}}\cdot ({\rm {\bf x}}-{\rm {\bf y}})}} ,
\end{equation}
and equating coefficients of like terms yields the coefficient commutation 
relations\cite{\textit{Quantum:1972}$^{,}$\cite{Jauch:1976}$^{,}$\cite{Gupta:1977}}
\begin{equation}
\label{eq11}
\left[ {\underline{a}(k),\underline{a}^\dag ({k}')} \right]=\left[ 
{\underline{b}(k),\underline{b}^\dag ({k}')} \right]=\,\,{\kern 1pt}{\kern 
1pt}-{\kern 1pt}\;\delta _{k{k}'} .
\end{equation}
Note the above relations differ from their traditional solution counterparts 
by the minus sign on the RHS\footnote{ Relation (\ref{eq11}) leads to an indefinite 
metric in the Fock space of states and negative norms for some state 
vectors. See Refs. \cite{Quantum:1972}, \cite{Jauch:1976}, and 
\cite{Gupta:1977}. Pauli\cite{Pauli:1941} delineated why operators 
with such a commutator could not correspond to real particle states. I do 
not take issue with Pauli on this point. I do suggest that his arguments may 
not apply to virtual particle states, and that the advantage shown by 
supplemental solutions in providing a possible answer to the vacuum energy 
problem was unknown in his time. This issue does, however, need to be 
resolved.} \cite{Pauli:1941}$^{,}$\cite{Dirac:1942}. This resulted from 
the time derivative in (\ref{eq9}), since the supplemental solutions have opposite 
signs from the traditional solutions for the time (energy) term in the 
exponent.

Using (\ref{eq6}) and (\ref{eq7}) in the relevant term in the Hamiltonian density
\begin{equation}
\label{eq12}
\underline{{\cal H}}_{\,\underline{\phi }}^0 =\sum\limits_r {\underline{\pi 
_r }} \underline{\mathop {\phi _r ^\dag }\limits^. }-\underline{{\cal 
L}}_{\,\underline{\phi }}^{\,0} =\underline{\mathop \phi \limits^. 
}\underline{\mathop {\phi ^\dag }\limits^. }+\underline{\mathop {\phi ^\dag 
}\limits^. }\underline{\mathop \phi \limits^. }-\underline{{\cal 
L}}_{\,\underline{\phi }}^{\,0} =\underline{\mathop \phi \limits^. 
}\underline{\mathop {\phi ^\dag }\limits^. }+\nabla \underline{\phi }^\dag 
\nabla \underline{\phi }+m^2\underline{\phi }^\dag \underline{\phi 
}\underline{}\underline{}
\end{equation}
and integrating over all space yields
\begin{equation}
\label{eq13}
\underline{H}=\sum\limits_{\rm {\bf k}} {\omega _{\rm {\bf k}} } \left\{ 
{\underline{a}^\dag (k)\underline{a}(k)\;\,-\,\;\textstyle{1 \over 
2}\;\,+\,\;\underline{b}^\dag (k)\underline{b}(k)\;\,-\,\;\textstyle{1 \over 
2}} \right\}
\end{equation}
where for notational streamlining we here and from henceforth drop the sub 
and superscript notation. Note the $\raise.5ex\hbox{$\scriptstyle 
1$}\kern-.1em/ \kern-.15em\lower.25ex\hbox{$\scriptstyle 2$} $ terms in (\ref{eq13}) 
have the opposite sign from similar terms in the traditional Hamiltonian
\begin{equation}
\label{eq14}
H=\sum\limits_{\rm {\bf k}} {\omega _{\rm {\bf k}} } \left\{ {a^\dag 
(k)a(k)\;\;+\;\,\textstyle{1 \over 2}\;\,+\,\;b^\dag 
(k)b(k)\;\;+\;\textstyle{1 \over 2}\;} \right\}.
\end{equation}
The presence of the $\raise.5ex\hbox{$\scriptstyle 1$}\kern-.1em/ 
\kern-.15em\lower.25ex\hbox{$\scriptstyle 2$} $ terms in the traditional 
development of QFT implied an infinite energy VEV, and necessitated the \textit{ad hoc} 
introduction of normal ordering in order to keep vacuum energy zero. 
However, if we consider the total (free, scalar) Hamiltonian as
\begin{equation}
\label{eq15}
H_{\mbox{tot}} =H+\underline{H}=\sum\limits_{\rm {\bf k}} {\omega _{\rm {\bf 
k}} } \left\{ {a^\dag (k)a(k)\;\;+\;\;b^\dag 
(k)b(k)\;\;+\;\;\underline{a}^\dag 
(k)\underline{a}(k)\;\;+\;\;\underline{b}^\dag (k)\underline{b}(k)} 
\right\}
\end{equation}
then the $\raise.5ex\hbox{$\scriptstyle 1$}\kern-.1em/ 
\kern-.15em\lower.25ex\hbox{$\scriptstyle 2$} $ terms all drop out and the 
expectation energy of the vacuum is naturally zero without having to resort 
to normal ordering\footnote{ Teller (Ref. \cite{Teller:1995}) discusses 
aspects of QFT that seem unnatural and inelegant, and he includes normal 
ordering of the Hamiltonian as one of these. He submits that a complete and 
true theory would not have such an artificial, and arbitrarily imposed, 
feature. On page 130, with reference to normal ordering he states, ``If, as 
appears to be the case, at this point one must use mathematically 
illegitimate tricks, concern is an appropriate response.'' It is noteworthy 
that nowhere else in QFT are we permitted to simply assume that 
non-commuting operators temporarily commute. This seeming contradiction is 
resolved by incorporating supplemental solutions into the 
theory.}\cite{Teller:1995}.

\subsection{Dirac and Pro\c{c}a Supplemental Solutions}
\label{subsec:dirac}
As should be expected, the above analysis has its analogues for spin 
$\raise.5ex\hbox{$\scriptstyle 1$}\kern-.1em/ 
\kern-.15em\lower.25ex\hbox{$\scriptstyle 2$} $ and spin 1 fields. The 
Pro\c{c}a equation is so closely related to the Klein-Gordon equation that 
all results of the preceding sections can be readily extrapolated to spin 1 
fields. 

It is not quite so obvious, however, that the results of the preceding 
sections can be extrapolated to Dirac particles. This is because the 
conjugate momentum for a Dirac field does not involve a time derivative of 
that field, and hence one cannot expect the Dirac equation solutions to 
directly parallel (\ref{eq9}) through (\ref{eq11}). Further, the Dirac Hamiltonian for the 
traditional solutions has negative terms of $\raise.5ex\hbox{$\scriptstyle 
1$}\kern-.1em/ \kern-.15em\lower.25ex\hbox{$\scriptstyle 2$} \omega 
_{k}$ analogous to the positive such terms in (\ref{eq14}).

In Appendix B, the coefficient anticommutaion relations for spin 
$\raise.5ex\hbox{$\scriptstyle 1$}\kern-.1em/ 
\kern-.15em\lower.25ex\hbox{$\scriptstyle 2$} $ supplemental particles are 
shown to have the opposite sign from their traditional solutions 
counterparts, i.e.,
\begin{equation}
\label{eq16}
\left[ {\underline{c}_r (p),\underline{c}_s^\dag ({p}')} \right]_+ =\left[ 
{\underline{d}_r (p),\underline{d}_s^\dag ({p}')} \right]_+ =\,\,{\kern 
1pt}{\kern 1pt}-{\kern 1pt}\;\delta _{p{p}'} \delta _{rs} .
\end{equation}
Using (\ref{eq16}) in the free Hamiltonian density for Dirac supplemental particles 
results in a total Dirac particle Hamiltonian analogous to (\ref{eq15}), i.e, having 
no $-\raise0.5ex\hbox{$\scriptstyle 
1$}\kern-0.1em/\kern-0.15em\lower0.25ex\hbox{$\scriptstyle 2$}$ terms.

\subsection{Supplemental Operators, Propagators, and Observables}
\label{subsec:supplemental}
Analogous results can be found for supplemental field number operators, 
propagators, and observables, and derivations of these will be shown in a 
subsequent article. Number operators yield number eigenvalues of opposite 
sign from their traditional counterparts. Supplemental propagators have the 
same form, but opposite sign from traditional propagators. Observables of 
states created and destroyed by supplemental operators, in general, do not 
correspond to those of the physically manifest world.

Of particular note, the total Hamiltonian expectation value for a 
supplemental particle state is negative. Further, the total three-momentum 
expectation value for a supplemental particle state is in the opposite 
direction of particle velocity. Such characteristics are not those of real 
particles, though they can be so for virtual particles. For example, the 
virtual exchange between two oppositely charged macroscopic bodies must 
entail three-momentum in the opposite direction of travel of the virtual 
particles in order for the bodies to attract. And virtual loop diagrams are 
integrated over both positive and negative energies for the individual 
virtual particles therein. Further, scalar (timelike polarization) virtual 
photons have negative energies.\cite{Mandl:1984} Still further, fermion 
zero-point energies, as noted above, are negative.

\subsection{Nature of Supplemental Particles}
\label{subsec:nature}
Hence, we postulate herein that if supplemental states are indeed realized 
in the spectrum of states, they are necessarily constrained to be virtual, 
cannot be real, and are never directly observed. No symmetry breaking 
mechanism (at least none much above contemporary energy levels, such as that 
conjectured between particles and sparticles in supersymmetry) is envisioned 
between the traditional and supplemental particles. And to be clear, the 
supplemental particles, though having negative energy states, are not a 
reincarnation of Dirac's ``sea of negative energy'', but quite a different 
thing entirely.

\section{Vacuum Energy and the Cosmological Constant}
\subsection{Cancellation of Zero-point Energy Fluctuations}
Weinberg\cite{Weinberg:1989}, among others, notes that summing of the 
zero-point energies ($\raise.5ex\hbox{$\scriptstyle 1$}\kern-.1em/ 
\kern-.15em\lower.25ex\hbox{$\scriptstyle 2$} $ quantum at each frequency) 
of all normal modes of some field of mass $m$ up to a wave number cutoff $k_c 
\gg m$ yields a vacuum energy density (with $\hbar =c =$1)
\begin{equation}
\label{eq17}
<\rho _0 >\;=\;\int_0^{k_c } {\frac{4\pi k^2dk}{(2\pi )^3}} \frac{1}{2}\sqrt 
{k^2+m^2} \;\cong \;\frac{k_c ^4}{16\pi ^2}.
\end{equation}
For a suitable cutoff on the order of the Planck scale, we get
\begin{equation}
\label{eq18}
<\rho _0 >\,\cong 2\times 10^{71}\,\,\mbox{GeV}^4,
\end{equation}
which, being off from the observed density of the vacuum
\begin{equation}
\label{eq19}
\rho _V \,\le \,10^{-47}\,\,\mbox{GeV}^4
\end{equation}
by a factor on the order of 118 decimal places, is the well-known biggest 
discrepancy between theory and experiment in the history of science.

Further, the zero point energy density from $\raise.5ex\hbox{$\scriptstyle 
1$}\kern-.1em/ \kern-.15em\lower.25ex\hbox{$\scriptstyle 2$} $ quanta normal 
modes $\left\langle {\rho _0 } \right\rangle $ predicted by traditional 
quantum field theory does not result in pressures $\left\langle {p_0 } 
\right\rangle =-\left\langle {\rho _0 } \right\rangle $. (See background 
review in Appendix C.) This implies that vacuum fluctuations cannot manifest 
as a cosmological constant, regardless of size. More importantly, the zero 
point stress-energy tensor from $\raise.5ex\hbox{$\scriptstyle 
1$}\kern-.1em/ \kern-.15em\lower.25ex\hbox{$\scriptstyle 2$} $ quanta normal 
modes $\left\langle {T_0^{\mu \nu } } \right\rangle $ is then not Lorentz 
invariant. Further, in an expanding universe, vacuum energy density would 
not be constant.

Consider, however, that inclusion of the supplemental solutions into the 
theory results in a complete cancellation of each positive energy vacuum 
normal mode by a negative energy vacuum normal mode of the precise same 
magnitude, and hence a null total zero point energy density VEV, i.e.,
\begin{equation}
\label{eq20}
<_{Total} \rho _0 >\,=\,<\rho _0 +\underline{\rho _0 }>\,=0,
\end{equation}
which is consistent with (\ref{eq19}) and constant over time. The stress-energy 
tensor $\left\langle {_{Total} T_0^{\mu \nu } } \right\rangle $ is also 
null, and thus obviously Lorentz invariant.

As discussed by Weinberg, Ellwenger\cite{Ellwanger:1}, Peebles and 
Ratra\cite{Peebles:1}, and others, some mechanisms investigated to null 
out the zero point energy, such as supersymmetry, work only to certain 
energy scales, resulting in enormous predicted values for our epoch. Only a 
mechanism that transcends symmetry breaking scales and is effective over 
virtually \textit{all} energy levels can possibly produce a null (or near null) $\rho 
_{V}$. Supplemental solutions provide such a scale invariant symmetric 
mechanism.

\subsection{Scalar Field Vacuum Potentials}
\label{subsec:scalar}
Higgs, inflaton, and quintessence theories posit scalar fields with 
potentials leading to a post symmetry breaking Lagrangian density for a real 
field of the form\cite{Peebles:1993}$^{,}$\cite{Ref:1}$^{,}$\cite{Ref:2}
\begin{equation}
\label{eq21}
{\cal L}_\phi =\textstyle{1 \over 2}\left( {\partial _\mu \phi g^{\mu \nu 
}\partial _\nu \phi -m^2\phi ^2} \right)\,-\,V,
\end{equation}
where $V$ is constant on the order of $m^{4}$. This leads to mass-energy 
density and pressure in local Minkowski coordinates of
\begin{equation}
\label{eq22}
T_\phi ^{00} =\rho _\phi =\textstyle{1 \over 2}\left( {\left( {\dot {\phi }} 
\right)^2+\left( {\nabla \phi } \right)^2+m^2\phi ^2} \right)+V,
\end{equation}
\begin{equation}
\label{eq23}
T_\phi ^{11} =p_\phi =\left( {\partial _1 \phi } \right)^2+\textstyle{1 
\over 2}\left( {\left( {\dot {\phi }} \right)^2-\left( {\nabla \phi } 
\right)^2-m^2\phi ^2} \right)-V.
\end{equation}
Einstein's field equations are
\begin{equation}
\label{eq24}
G^{\mu \nu }-\Lambda g^{\mu \nu }=8\pi GT^{\mu \nu }+8\pi G\left\langle 
{T^{\mu \nu }} \right\rangle .
\end{equation}
where $\Lambda $ is the classical cosmological constant. From the results of 
Appendix C, (\ref{eq22}), and (\ref{eq23}), one can see that
\begin{equation}
\label{eq25}
T_\phi ^{00} =\frac{1}{V_s }\sum\limits_{\rm {\bf k}} {\omega _{\rm {\bf k}} 
} \left\{ {a^\dag (k)a(k)\;\;+\;\,\textstyle{1 \over 2}} \right\}+V
\end{equation}
\begin{equation}
\label{eq26}
T_\phi ^{11} =\frac{1}{V_s }\sum\limits_{\rm {\bf k}} {\frac{\left( {k_1 } 
\right)^2}{\omega _{\rm {\bf k}} }} \left\{ {a^\dag 
(k)a(k)\;\;+\;\,\textstyle{1 \over 2}} \right\}-V,
\end{equation}
where $V_{s}$ is spatial volume. Thus, the vacuum expectation values become
\begin{equation}
\label{eq27}
\left\langle {T_\phi ^{00} } \right\rangle \,=\,\left\langle {\rho _\phi } 
\right\rangle =\frac{1}{V_s }\sum\limits_{\rm {\bf k}} {\frac{\omega _{\rm 
{\bf k}} }{2}} +V\,\,\,\,\,\,\,\,\,\,\,\,\,\,\,\,\,\,\,\left\langle {T_\phi 
^{11} } \right\rangle \,=\,\left\langle {p_\phi } \right\rangle 
=\frac{1}{V_s }\sum\limits_{\rm {\bf k}} {\frac{\left( {k_1 } 
\right)^2}{2\omega _{\rm {\bf k}} }} -V.
\end{equation}
If one ignores the summations, then from (\ref{eq24}), the constant potential $V$ acts 
like a contribution to the effective cosmological constant
\begin{equation}
\label{eq28}
\Lambda _{\mbox{eff}} =\Lambda +\Lambda _\phi =\Lambda +8\pi G\left\langle 
{\rho _\phi } \right\rangle =\Lambda +8\pi GV
\end{equation}
(or alternatively, like the vacuum density of the scalar field $\left\langle 
{\rho _\phi } \right\rangle =-\left\langle {p_\phi } \right\rangle $, where 
pressure is negative.) 

Two significant problems exist with this approach.

\begin{enumerate}
\item The calculated order of $\Lambda _{\phi }$ differs from observation by many orders of magnitude. For electroweak symmetry breaking, $V$ is about 10$^{8}$ GeV$^{4}$. (Compare with (\ref{eq19})) For GUT symmetry breaking, $V$ is on the order of 10$^{64}$ GeV$^{4}$.
\item There is no justification for ignoring the $\raise.5ex\hbox{$\scriptstyle 1$}\kern-.1em/ \kern-.15em\lower.25ex\hbox{$\scriptstyle 2$} $ quanta normal vacuum modes summation in the VEV's of (\ref{eq27}), which as noted earlier, are not only immense, but destroy vacuum Lorentz invariance and constancy.
\end{enumerate}
However, if supplemental particles $\underline{\phi }$ are included in the 
calculations, the vacuum expectation stress energy tensor should be null. 
The vacuum fluctuation summations then have two components, equal in 
magnitude and opposite in sign, that cancel. Additionally, given the form 
for the supplemental particle Hamiltonian (\ref{eq13}), it is reasonable to expect 
the supplemental scalar field ${\phi }$ Lagrangian to have, due to symmetry, 
the similar functional dependence as (\ref{eq21}), but with opposite sign for the 
constant potential energy density\footnote{ Linde (Refs 
\cite{Linde:1984}, \cite{Linde:1988}) also suggests negative energy 
particles in order to obtain a null cosmological constant, but, unlike the 
present approach, starts by augmenting the Lagrangian with a term equal to 
the original Lagrangian but having opposite sign. Linde's approach and the 
one shown herein have some similarities, but are quite different in other 
regards. One of the similarities is the positing of an additional scalar 
potential with opposite sign, such as suggested in this section. Certain 
possible results from adopting this approach that are discussed in this and 
the subsequent two sections were also noted by Linde} 
\cite{Linde:1984}$^{,}$\cite{Linde:1988}, i.e.,
\begin{equation}
\label{eq29}
\underline{{\cal L}}_{\underline{\phi }} =\textstyle{1 \over 2}\left( 
{\partial _\mu \underline{\phi }g^{\mu \nu }\partial _\nu \underline{\phi 
}-m^2\underline{\phi }^2} \right)\,\,-\,\,\underline{V}=\textstyle{1 \over 
2}\left( {\partial _\mu \underline{\phi }g^{\mu \nu }\partial _\nu 
\underline{\phi }-m^2\underline{\phi }^2} \right)\,\,+\,\,V.
\end{equation}
With this one gets
\begin{equation}
\label{eq30}
\underline{T_\phi ^{00} }=\underline{\rho _\phi }=\textstyle{1 \over 
2}\left( {\left( {\underline{\dot {\phi }}} \right)^2+\left( {\nabla 
\underline{\phi }} \right)^2+m^2\underline{\phi }^2} \right)-V
\end{equation}
\begin{equation}
\label{eq31}
\underline{T_\phi ^{11} }=\underline{p_\phi }=\left( {\partial _1 
\underline{\phi }} \right)^2+\textstyle{1 \over 2}\left( {\left( 
{\underline{\dot {\phi }}} \right)^2-\left( {\nabla \underline{\phi }} 
\right)^2-m^2\underline{\phi }^2} \right)+V,
\end{equation}
and
\begin{equation}
\label{eq32}
\left\langle {_{Tot} T_\phi ^{00} } \right\rangle \,=\,\left\langle {\rho 
_\phi +\underline{\rho _\phi }} \right\rangle =\underbrace {\sum\limits_{\rm 
{\bf k}} {\textstyle{1 \over {2V_s }}\left( {\omega _{\rm {\bf k}} -\omega 
_{\rm {\bf k}} } \right)} 
}_{\mbox{from}\,\mbox{zero}\,\mbox{point}\,\mbox{fluctuations}}+\underbrace 
{V-V}_{\begin{array}{l}
 \mbox{from}\,\mbox{scalar} \\ 
 \mbox{potentials} \\ 
 \end{array}}\,=0
\end{equation}
\begin{equation}
\label{eq33}
\left\langle {_{Tot} T_\phi ^{11} } \right\rangle \,=\,\left\langle {p_\phi 
+\underline{p_\phi }} \right\rangle =\underbrace {\sum\limits_{\rm {\bf k}} 
{\textstyle{1 \over {2V_s }}\left( {\frac{\left( {k_1 } \right)^2-\left( 
{k_1 } \right)^2}{\omega _{\rm {\bf k}} }} \right)} 
}_{\mbox{from}\,\mbox{zero}\,\mbox{point}\,\mbox{fluctuations}}-\underbrace 
{V+V}_{\begin{array}{l}
 \mbox{from}\,\mbox{scalar} \\ 
 \mbox{potentials} \\ 
 \end{array}}\,=0.
\end{equation}
The zero point fluctuation summations from the supplemental and traditional 
particles, as well as the post symmetry breaking level potentials, cancel, 
leaving a Lorentz invariant, constant, null stress-energy tensor for the 
vacuum and no cosmological constant.

\subsection{Small Cosmological Constant}
\label{subsec:small}
Consider a slight asymmetry between one scalar field and the other with 
regard to their constant vacuum potentials, wherein
\begin{equation}
\label{eq34}
\underline{{\cal L}}_{\underline{\phi }} =\textstyle{1 \over 2}\left( 
{\partial _\mu \underline{\phi }g^{\mu \nu }\partial _\nu \underline{\phi 
}-m^2\underline{\phi }^2} \right)\,\,+\,\,\underline{V},
\end{equation}
such that
\begin{equation}
\label{eq35}
V=\underline{V}+\delta ,
\end{equation}
with $\delta $ small. Using this in (\ref{eq32}) and (\ref{eq33}), we would have
\begin{equation}
\label{eq36}
\left\langle {_{Tot} T_\phi ^{\mu \nu } } \right\rangle \,=\,\delta \left[ 
{{\begin{array}{*{20}c}
 1 \hfill & 0 \hfill & 0 \hfill & 0 \hfill \\
 0 \hfill & {-1} \hfill & 0 \hfill & 0 \hfill \\
 0 \hfill & 0 \hfill & {-1} \hfill & 0 \hfill \\
 0 \hfill & 0 \hfill & 0 \hfill & {-1} \hfill \\
\end{array} }} \right],
\end{equation}
yielding a small cosmological constant, as well as a Lorentz invariant 
stress-energy tensor for the vacuum with constant vacuum energy density.

This is as we see it today, although we still need a large cosmological 
constant in earlier epochs for inflation to be viable.

\subsection{Inflation}
Possibilities exist for time dependent $\delta $. A large $\delta $ at 
earlier times that relaxed to a present epoch smaller value would be one way 
to satisfy inflationary requirements.

In inflation scenarios (prior to, and concurrent with, a symmetry breaking 
process), the scalar field Lagrangian density is typically expressed as
\begin{equation}
\label{eq37}
{\cal L}_\phi =\textstyle{1 \over 2}\partial _\mu \phi g^{\mu \nu }\partial 
_\nu \phi \,-\,V(\phi )
\end{equation}
where $V$(\textit{$\phi $}) can have a \textit{$\phi $} dependence as depicted in the top half of Figure 1. 
Inflation begins when the energy density of the \textit{$\phi $} field is concentrated in a 
high, relatively flat potential (the false vacuum at time $t_{1}$ in the 
figure) and ends with the \textit{$\phi $} field in its lowest potential state (the true 
vacuum at time $t_{2})$. The remnant value $V$ then becomes part of the post 
symmetry breaking Lagrangian (see (\ref{eq21})\footnote{ The $\phi $ field in (\ref{eq21}) 
is different from the $\phi $ field in (\ref{eq37}), however. The former represents 
a ``new'' particle type whose zero real density state corresponds to the 
true vacuum of time $t_{2}$ in Figure 1. The latter represents a different 
(though closely related) particle whose zero real density state corresponds 
to the false vacuum of time $t_{1}$ in Figure 1.}).

Consider, on the bottom part of Figure 1, a supplemental particle potential, 
which could be expected to be a mirror image of the traditional potential of 
the left side. The particles we see in our universe, including all those 
coupled to the \textit{$\phi $} field, would have their destinies tied to the potential 
$V$(\textit{$\phi $}), and presumably not to $\underline{V}(\underline{\phi })$. Yet, the 
vacuum, and its properties, would be tied to both (the sum of the two.) 

A range of possibilities exist with regard to the temporal relationship 
between the two sides of Figure 1. Consider one extreme in which both 
potentials (as they appear in the figure) are precise mirrors of each other, 
and both the \textit{$\phi $} and $\underline{\phi }$ fields move along their respective 
potential curves in lockstep. At each moment in time, the expectation values 
of each field are located at the same point along their respective 
horizontal axes. At no time would we have an inflationary occurrence, since 
the total vacuum potential, and thus the vacuum stress-energy tensor, would 
always be zero.

To get inflation, at least one of two possibilities need exist. First, the 
potentials might, for some reason, not be precise mirror images of one 
another. This could result in a large total potential (due to the sum of the 
two field potentials) at early times, but a very small total potential in 
the present (giving rise to a small cosmological constant.) This would, of 
course, also give rise to different time evolutions for the expectation 
values of the two fields.

\begin{figure}
\centerline{\includegraphics{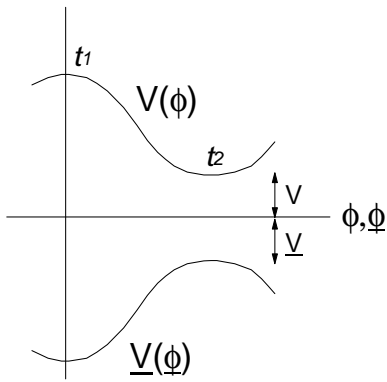}}
\caption{. Possible Traditional and Supplemental Scalar Field Potentials}
\label{fig1}
\end{figure}

Second, the time evolution along the two curves might differ, even if they 
have the same (but inverted) shapes. The supplemental potential vacuum, 
mirroring its traditional counterpart, would tend to seek the state of 
\textit{highest} potential (lowest magnitude potential). With the real particles of the 
universe having initial high positive energy, perhaps the impetus for the 
supplemental field to seek its (inverted) minimum would be stronger than 
that of the traditional field. Thus, the supplemental field might reach its 
expectation value for the true vacuum while the traditional field was still 
at the false vacuum. The net potential from the sum of the two potentials at 
that time would have a large positive value and initiate inflation, which 
would continue until the traditional field reached true vacuum. If any 
supplemental particles exist at that time, during inflation they would be 
thinned out to effectively zero density, leaving a net positive energy 
particle density in the universe.

A small residual potential might exist if the last part of the journey takes 
an extended time, so a small discrepancy in magnitude difference between the 
two fields still remains. In time, this residual could be expected to 
diminish to zero.

\section{Antipodal Symmetry and Supplemental Solutions}
Linde\cite{Linde:1988} conjectured ``some mechanism, associated probably 
with some kind of symmetry of the elementary particle theory, which would 
automatically lead to a vanishing of the cosmological constant in a wide 
class of theories''. He considered the possibility of two quasi-independent 
``universes'' with the same particle equations of motion, but with opposite 
signs for the Lagrangian, and thus for energy, for which he coined the term 
``antipodal symmetry''. One universe (ours) has positive energy particles. 
The other is a mirror image with negative energy particles. The total 
energy, and the total cosmological constant, net to zero.

As shown herein, such a symmetry already exists within the mathematics of 
quantum field theory. Though largely unrecognized, mathematical solutions of 
the extant theory provide the desired symmetry. Nothing more is needed. The 
solution is direct, complete, and remarkably simple. The fundamental 
question then is whether or not nature in her physical manifestation 
parallels the nature of her mathematics. Do virtual supplemental particles 
actually exist?

A partial answer, as noted in Section \ref{subsec:supplemental}, is that 
traditional QFT calculations already encompass (half of the time) virtual 
particles with negative energy and other characteristics of supplemental 
particles. It is not a big step to consider vacuum fluctuations and scalar 
field potentials following suit.

The absence of observed real supplemental particles may be conjectured as 
being related to the fact that reversing the arrow of time in the 
supplemental solutions produces the traditional solutions. That may mean, 
with regard to perception, that real supplemental particles travel backward 
in time. Thus one might speculate that the universe needed no initial energy 
from which to begin, with equal amounts of traditional and supplemental 
particles emerging from nothing. From there, the real negative energy 
supplemental particles traveled backward in time, and in the process created 
their own universe, which would appear to any beings in that universe to 
possess positive energy. The traditional particles, traveling forward in 
time, created our universe and appear to us as having positive energy. Each 
universe would have only one kind of real particle, but virtual particles of 
both types. And this would result in null vacuum energy and null (or near 
null) cosmological constants in both\footnote{ Linde\cite{Linde:1988} 
cautions that ``.. it is dangerous to consider theories containing particles 
\textit{with both signs of energy}, which typically occurs in the theories containing particles with 
indefinite metric..'' but notes that ``..in our case no such problem does 
exist [because] the [traditional] fields do not interact with the [negative 
energy] fields..'' In the present suggested scenario, real traditional 
particles would not interact with real supplemental particles, because the 
latter would not be around. Alternatively, the traditional and supplemental 
particles could be completely decoupled, yet each would affect the 
properties of spacetime in opposite ways.}.

In this scenario, the act of creation itself would have been the first, and 
most fundamental, symmetry breaking of nature's laws.

\section{Summary}
A previously unrecognized, but fundamental, symmetry in elementary particle 
theory exists in which supplemental solutions to the QFT field equations are 
obtained from the traditional solutions by taking $\omega \to -\omega $. The 
implementation of this symmetry in the theory results in a total Hamiltonian 
yielding a null energy VEV, renders the \textit{ad hoc} procedure of normal ordering 
unnecessary, and predicts a null cosmological constant. An asymmetry between 
the traditional and supplemental particle solutions could lead to a non-null 
cosmological constant, whose magnitude and evolution would depend on the 
nature and degree of the asymmetry. Agreement with observations is only 
maintained if supplemental states occur only as virtual, and not as real, 
particles, and given the properties of the supplemental states, this appears 
reasonable.

The supplemental/traditional solution symmetry can be maintained over all 
energy scales, unlike other attempts to null out vacuum energy, such as 
supersymmetry, which only succeed down to particular, non-contemporary, 
energy levels. This symmetry provides a simple, direct solution to the 
predicted enormous vacuum energy problem, and that solution is already built 
into the original theory. No new theory, and no \textit{ad hoc} complications are required. 

\section{Acknowledgments}
I would like to extend my thanks and appreciation to Stephen Kelley, Andrei 
Linde, and Robin Ticciati, for reading early versions of the manuscript, and 
for offering valuable insights, suggestions, and references.

\section{Appendix A: Supplemental Solutions Independence from the Traditional 
Solutions}
Since the traditional solutions of the Klein-Gordon equation
\begin{equation}
\label{eq38}
\phi \underline{}=\phi ^+\underline{}+\phi ^-=\sum\limits_{\rm {\bf k}} 
{\left( {\frac{1}{2V\omega _k }} \right)^{1/2}\left\{ {a({\rm {\bf 
k}})e^{ikx}+b^\dag ({\rm {\bf k}})e^{-ikx}} \right\}} 
\end{equation}
are summed over all \textbf{k}, for every $k_{x}$ = 10, for example, in (\ref{eq38}), 
a $k_{x}$ = --10 is also summed. Thus, some might reason, for each +$E$ in the 
summation with a --$k_{x}$, there is a +$E$ with with a +$k_{x}$, and similarly 
for --$E$, leading to the solution forms (\ref{eq4}) and (\ref{eq5}). There are, however, 
subtle differences between sign in an algebraic relation and sign of an 
algebraic quantity, which is illustrated with two elementary examples as 
follows.

Example 1

Consider the difference between sign in an algebraic expression and sign of 
an algebraic quantity in that expression. For example,
\begin{equation}
\label{eq39}
x-y=0
\end{equation}
has fully symmetric values for y. For every positive value of y, there is a 
negative value of same magnitude. But that does not mean that the relation
\begin{equation}
\label{eq40}
x+y=0
\end{equation}
is contained within (\ref{eq39}). Precisely parallel logic applies to supplemental 
vs traditional solutions.

Example 2

For ease of illustration, consider the K-G solutions as states, as in 
relativistic quantum mechanics. Take the expression for constant phase for a 
wave traveling along the x axis 
\begin{equation}
\label{eq41}
\omega t-k_x x=\,\mbox{constant},
\end{equation}
differentiate with respect to $t$, and solve for the wave velocity.
\begin{equation}
\label{eq42}
v\mbox{ }=\mbox{ }dx/dt\mbox{ }=\mbox{ }\omega /k_x 
\end{equation}
and velocity has the same sign as $k_{x}$ (momentum and velocity are in same 
direction.) For $k_{x}$ having opposite sign, $v$ also has opposite sign.

Repeat for the supplemental solutions.
\begin{equation}
\label{eq43}
\omega t+k_x x=\,\mbox{constant}
\end{equation}
and
\begin{equation}
\label{eq44}
v\mbox{ }=\mbox{ }dx/dt\mbox{ }=\mbox{ -}\omega /k_x .
\end{equation}
So whatever the sign on the 3 momentum $k_{x}$, the velocity of the wave is 
in the opposite direction. Change the sign of $k_{x}$, and the velocity is 
still in the opposite direction of $k_{x}$ (it changes sign too).

While it may seem bizarre to have velocity and 3-momentum in opposite 
directions, there is no doubt that it is a radically different result, with 
radically different physical implications. In fact, it is so different that 
no real, physical particles/waves display it. When summing over all positive 
and negative values of $k_{x}$ in the traditional general solution for $\phi 
$, one does NOT get this result for the negative $k_{x}$ values (velocity is 
still in the direction of $k_{x}$ for negative $k_{x})$.

As noted in Section \ref{subsec:supplemental}, the same thing is true in 
quantum field theory, where the solutions are operator fields, rather than 
states.

\section{Appendix B: Supplemental Dirac Anti-commutators}
\label{sec:appendix}
To derive the coefficient anti-commutator relations of (\ref{eq16}), begin with the 
Dirac equation for one of the supplemental particle solutions,
\begin{equation}
\label{eq45}
(i\gamma ^\mu \partial _\mu -m)\underline{\psi }^+=0.
\end{equation}
Insert
\begin{equation}
\label{eq46}
\underline{\psi }^+=\underline{u}({\rm {\bf p}})e^{i\underline{px}},
\end{equation}
choose a representation for gamma matrices such as the standard rep, and 
solve the resulting eignenvalue problem for \underline {\textit{u}}$^{(i)}$, 
where $i $= eigenvector number (1,2,3,4). Each eigenvector has four (implicit 
and unlabeled here) components in the representation space.

If, in parallel with the traditional approach, one then defines a normalized 
vector \underline {\textit{u}}$_{r}$ as
\begin{equation}
\label{eq47}
\underline{u}_r =(\textstyle{{-\;\vert E\vert +m} \over 
{2m}})^{1/2}\underline{u}^{(r)},
\end{equation}
we find
\begin{equation}
\label{eq48}
\underline{u}_r^\dag ({\rm {\bf p}})\underline{u}_s ({\rm {\bf 
p}})=\textstyle{{-\;\vert E\vert } \over m}\delta _{rs} ,
\end{equation}
which has the opposite sign on the RHS from the similar traditional 
relationship. Repeating the procedure for \underline {\textit{$\psi $}}$^{ 
-}$ yields the same relation for \underline {\textit{v}}$_{r}$(\textbf{p}).

Normalizing (\ref{eq46}) according to (\ref{eq47}), applying second quantization (with 
anti-commuting fields) to the supplemental Dirac solutions, and using (\ref{eq48}) 
in parallel fashion to steps (\ref{eq9}) through (\ref{eq11}), one ends up as promised with 
(\ref{eq16}), the coefficient anti-commutation relations for supplemental Dirac 
particles.

\section{Appendix C: Real Scalar Field Vacuum Stress Energy Tensor}
\label{sec:mylabel2}
From the free real scalar field Lagrangian density
\begin{equation}
\label{eq49}
{\cal L}_\phi =\textstyle{1 \over 2}\left( {\partial _\mu \phi g^{\mu \nu 
}\partial _\nu \phi -m^2\phi ^2} \right)
\end{equation}
it can be shown\cite{Ref:3} that
\begin{equation}
\label{eq50}
T_\phi ^{00} =\rho _\phi =\textstyle{1 \over 2}\left( {\left( {\dot {\phi }} 
\right)^2+\left( {\nabla \phi } \right)^2+m^2\phi ^2} \right)
\end{equation}
\begin{equation}
\label{eq51}
T_\phi ^{11} =p_\phi =\left( {\partial _1 \phi } \right)^2+\textstyle{1 
\over 2}\left( {\left( {\dot {\phi }} \right)^2-\left( {\nabla \phi } 
\right)^2-m^2\phi ^2} \right).
\end{equation}
The real scalar field and its derivatives are
\begin{equation}
\label{eq52}
\phi =\sum\limits_{\rm {\bf k}} {\frac{1}{\sqrt {2V_s \omega _{\rm {\bf k}} 
} }\left( {a({\rm {\bf k}})e^{-ikx}+a^\dag ({\rm {\bf k}})e^{ikx}} \right)} 
\end{equation}
\begin{equation}
\label{eq53}
\dot {\phi }=\sum\limits_{\rm {\bf k}} {\frac{i\omega _{\rm {\bf k}} }{\sqrt 
{2V_s \omega _{\rm {\bf k}} } }\left( {-a({\rm {\bf k}})e^{-ikx}+a^\dag 
({\rm {\bf k}})e^{ikx}} \right)} 
\end{equation}
\begin{equation}
\label{eq54}
\phi ,_i =\sum\limits_{\rm {\bf k}} {\frac{ik_i }{\sqrt {2V_s \omega _{\rm 
{\bf k}} } }\left( {a({\rm {\bf k}})e^{-ikx}-a^\dag ({\rm {\bf k}})e^{ikx}} 
\right)} ,
\end{equation}
where $V_{s}$ is spatial volume. The first term on the RH of (\ref{eq50}) is
\begin{equation}
\label{eq55}
\textstyle{1 \over 2}\dot {\phi }\dot {\phi }=\textstyle{1 \over 2}\left( 
{\sum\limits_{\rm {\bf k}} {\frac{i\omega _{\rm {\bf k}} }{\sqrt {2V_s 
\omega _{\rm {\bf k}} } }\left( {-a({\rm {\bf k}})e^{-ikx}+a^\dag ({\rm {\bf 
k}})e^{ikx}} \right)} } \right)\left( {\sum\limits_{{\rm {\bf {k}'}}} 
{\frac{i\omega _{{\rm {\bf {k}'}}} }{\sqrt {2V_s \omega _{{\rm {\bf {k}'}}} 
} }\left( {-a({\rm {\bf {k}'}})e^{-i{k}'x}+a^\dag ({\rm {\bf 
{k}'}})e^{i{k}'x}} \right)} } \right).
\end{equation}
In
\begin{equation}
\label{eq56}
\langle \phi _{\rm {\bf k}} \vert \rho _\phi \vert \phi _{\rm {\bf k}} 
\rangle 
\end{equation}
where $\rho _\phi $ is an operator represented by (\ref{eq50}), all terms with ${\rm 
{\bf k}}\ne {\rm {\bf {k}'}}$, drop out, as do all terms in $a({\rm {\bf 
k}})a({\rm {\bf k}})$ and $a^\dag ({\rm {\bf k}})a^\dag ({\rm {\bf k}})$. 
(\ref{eq55}) is part of (\ref{eq50}), so that part reduces to
\begin{equation}
\label{eq57}
\textstyle{1 \over 2}\dot {\phi }\dot {\phi }\to \textstyle{1 \over 
2}\sum\limits_{\rm {\bf k}} {\frac{\left( {\omega _{\rm {\bf k}} } 
\right)^2}{2V_s \omega _{\rm {\bf k}} }\left( {a({\rm {\bf k}})a^\dag ({\rm 
{\bf k}})+a^\dag ({\rm {\bf k}})a({\rm {\bf k}})} \right)} =\frac{1}{2V_s 
}\sum\limits_{\rm {\bf k}} {\omega _{\rm {\bf k}} \left( {a^\dag ({\rm {\bf 
k}})a({\rm {\bf k}})+\textstyle{1 \over 2}} \right)} ,
\end{equation}
where we have used the commutation relations in the last step. Similarly,
\begin{equation}
\label{eq58}
\textstyle{1 \over 2}\left( {\nabla \phi } \right)^2\to \frac{1}{2V_s 
}\sum\limits_{\rm {\bf k}} {\frac{\vert {\rm {\bf k}}\vert ^2}{\omega _{\rm 
{\bf k}} }\left( {a^\dag ({\rm {\bf k}})a({\rm {\bf k}})+\textstyle{1 \over 
2}} \right)} 
\end{equation}
\begin{equation}
\label{eq59}
\textstyle{1 \over 2}m^2\phi ^2\to \frac{1}{2V_s }\sum\limits_{\rm {\bf k}} 
{\frac{m^2}{\omega _{\rm {\bf k}} }\left( {a^\dag ({\rm {\bf k}})a({\rm {\bf 
k}})+\textstyle{1 \over 2}} \right)} .
\end{equation}
Since $\left( {\omega _{\rm {\bf k}} } \right)^2=m^2+\vert {\rm {\bf 
k}}\vert ^2$, the above three relations summed reduce to the well-known 
operator form for (\ref{eq50})
\begin{equation}
\label{eq60}
\rho _\phi =\frac{1}{V_s }\sum\limits_{\rm {\bf k}} {\omega _{\rm {\bf k}} 
\left( {a^\dag ({\rm {\bf k}})a({\rm {\bf k}})+\textstyle{1 \over 2}} 
\right)} ,
\end{equation}
which has the non-zero VEV $<\rho _\phi >$ equal to the sum of the $\omega 
_{k}$/2 terms.

In similar fashion, the pressure operator of (\ref{eq51}) reduces to
\begin{equation}
\label{eq61}
p_{1\phi } =\frac{1}{V_s }\sum\limits_{\rm {\bf k}} {\frac{\vert k_1 \vert 
^2}{\omega _{\rm {\bf k}} }\left( {a^\dag ({\rm {\bf k}})a({\rm {\bf 
k}})+\textstyle{1 \over 2}} \right)} 
\end{equation}
and therefore from (\ref{eq60}) and (\ref{eq61})
\begin{equation}
\label{eq62}
\langle \rho _\phi \rangle \ne -\langle p_{1\phi } \rangle .
\end{equation}
Thus, the vacuum fluctuations contribution to the stress energy tensor 
cannot result in a cosmological constant, which must have the form (in local 
Minkowskian coordinates)
\begin{equation}
\label{eq63}
\Lambda _V \eta ^{\mu \nu }=8\pi G\left[ {{\begin{array}{*{20}c}
 {\langle \rho _\phi \rangle } \hfill & 0 \hfill & 0 \hfill & 0 \hfill \\
 0 \hfill & {-\langle \rho _\phi \rangle } \hfill & 0 \hfill & 0 \hfill \\
 0 \hfill & 0 \hfill & {-\langle \rho _\phi \rangle } \hfill & 0 \hfill \\
 0 \hfill & 0 \hfill & 0 \hfill & {-\langle \rho _\phi \rangle } \hfill \\
\end{array} }} \right].
\end{equation}
Neither can the vacuum stress-energy tensor be Lorentz invariant, as only a 
tensor of the form $C\eta ^{\mu \nu }$, where $C$ is a constant, can be so.

It is well known\cite{Ref:4}, and derivable from the Einstein field 
equations with a Friedmann metric, that the rate of change of density in the 
universe is
\begin{equation}
\label{eq64}
\dot {\rho }=-3\left( {\rho +p} \right)\frac{\dot {a}}{a},
\end{equation}
where $a$ is the time dependent scale factor of the universe (the ``radius'' of 
the universe). So from (\ref{eq64}), the mass-energy density of the vacuum in an 
expanding universe can only remain constant if the vacuum pressure is equal 
in magnitude and opposite in sign from the vacuum density. As shown above in 
(\ref{eq62}), this is not the case for traditional fields vacuum fluctuation energy 
density.

\end{document}